\documentclass[PRL,aps,preprint,showpacs,amsmath,amssymb, superscriptaddress]{revtex4}

\usepackage{graphicx}
\usepackage{epsfig}
\usepackage{amsmath}
\usepackage{color}

\usepackage{bm}

\begin{document}

\title{Magnetic hierarchical deposition}
\author{Anna I. Posazhennikova}
\affiliation{Department of Physics, Royal Holloway, University of London,
  Egham, Surrey TW20 0EX, United Kingdom}
\author{Joseph O. Indekeu}
\affiliation{Institute for Theoretical Physics, KU Leuven, BE-3001 Leuven, Belgium}

\date{\today}

\begin{abstract}

We consider random deposition of debris or blocks on a line, with block sizes following a rigorous hierarchy: the linear size equals $1/\lambda^n$ in generation $n$, in terms of a rescaling factor $\lambda$. Without interactions between the blocks, this model is described by a logarithmic fractal, studied previously, which is characterized by a {\it constant} increment of the length, area or volume upon proliferation.  We study to what extent the logarithmic fractality survives, if each block is equipped with an Ising (pseudo-)spin $s=\pm 1$ and the interactions between those spins are switched on (ranging from antiferromagnetic to ferromagnetic). It turns out that the dependence of the surface topology on the interaction sign and strength is not trivial. For instance, deep in the ferromagnetic regime, our numerical experiments and analytical results reveal a sharp crossover from a Euclidean transient, consisting of aggregated domains of aligned spins, to an asymptotic logarithmic fractal growth. In contrast, deep into the antiferromagnetic regime the surface roughness is important and is shown analytically to be controlled by vacancies induced by frustrated spins. Finally, in the weak interaction regime, we demonstrate that the non-interacting model is extremal in the sense that the effect of the introduction of interactions is only quadratic in the magnetic coupling strength. In all regimes, we demonstrate the adequacy of a mean-field approximation whenever vacancies are rare. In sum, the logarithmic fractal character is robust with respect to the introduction of spatial correlations in the hierarchical deposition process.

\end{abstract}


\maketitle

\section{Introduction}
Aggregation and deposition phenomena have caught many a scientist's interest since their quantitative physical characterization following the introduction of the famous models of Eden growth and diffusion-limited aggregation (DLA) \cite{Eden,DLA}. Our goal in this paper is to merge, along the lines of the magnetic Eden model  and variants thereof \cite{MEM}, magnetic degrees of freedom and deposition rules. Such models, including the one we propose here, appeal to a multitude of physical circumstances in which order or disorder emerges subject to a local optimization criterion. The paradigm of this behavior is the selection of a configuration according to the minimization of the energy within a given interaction range around the probe degree of freedom. Many interesting works have appeared, in which the interplay of surface growth, surface roughness and critical phenomena near to or far from equilibrium have been studied, and useful insights have been gathered, complementing our understanding of cooperative phenomena at surfaces and interfaces \cite{KM,Many}.

Within the arena of surface growth models, the concept of fractality is omnipresent. Our second main goal in this paper is to develop a magnetic deposition model on an extraordinary class of fractal surfaces, physically significant but sparsely explored so far. The fractal media we have in mind are the so called logarithmic fractals \cite{logfrac}, characterized by an additive, rather than multiplicative, geometrical proliferation rule. While ordinary fractality refers to the proliferation of detail upon magnification in such a manner that the length, area, or volume, increases by a constant factor, logarithmic fractality implies an increment in the form of a constant term. Several works have sketched physical realizations and explored consequences of this unusual, but interesting class of models, for example, in the context of hierarchical deposition \cite{several} or bacterial biofilms \cite{bio}. We now turn to the assembly of these concepts with magnetic degrees of freedom in an Ising model spirit and formulate the main question of our research. We ask whether the logarithmic fractal character of the hierarchical deposition landscape is robust with respect to the introduction of magnetic interactions.

\section{Model}

We start from the hierarchical deposition model introduced in \cite{logfrac}, and assume a one-dimensional surface on which debris is deposited. In the absence of magnetic interactions a block is deposited with probability $P$. The block size follows a strict hierarchy. The linear size of the block is $\lambda^{-n}$ in generation $n$, and $\lambda^{n}$ sequential attempts are made to deposit blocks along the unit line $[0,1]$. Without loss of generality we assume a rescaling factor $\lambda =3$.   In the presence of magnetic interactions each block carries an Ising spin $s = \pm 1$ and the Hamiltonian we employ is of Blume-Emery-Griffiths type \cite{BEG} (however, without biquadratic interaction)
\begin{eqnarray}
H =  -J\sum_{<ij>} \sigma_i\sigma_j -\mu  \sum_{i}\sigma_i^2.
\label{BEG}
\end{eqnarray}
Here, $<ij>$ denotes nearest neighbours, $J$ is an exchange energy ($>0$ for ferro- and $<0$ for antiferromagnetic interactions) and $\mu$ is a chemical potential for deposited blocks. Blocks are nearest neighbours if they have one edge or part of one edge in common (sharing a corner is not enough). Further, the substrate on which the blocks are deposited may or may not be endowed with a spin. For our study we assume a {\em non-magnetic substrate}. Note that the allowed spin values ($\pm 1$ or zero) are independent of the sizes of the blocks. The reduced Hamiltonian reads
\begin{eqnarray}
-\beta H =  K\sum_{<ij>} \sigma_i\sigma_j + \Phi  \sum_{i}\sigma_i^2,
\label{RBEG}
\end{eqnarray}
with $\beta = 1/k_BT$ ($k_B$ is the Boltzmann constant and $T$ the absolute temperature), $K= J/k_BT$ and $\Phi = \mu/k_BT$.

Due to the magnetic interactions the deposition probability is not homogeneous but depends on the local field originating from already deposited block spins and on the chemical potential. Consequently, the probability for depositing an ``up" block in generation $n$, is
\begin{eqnarray}
P_i(+) = \exp(K\sum_{<ij>} s_j + \Phi) /Z_i,
\label{Pplus}
\end{eqnarray}
and that for depositing a ``down" block
\begin{eqnarray}
P_i(-) = \exp(-K\sum_{<ij>} s_j + \Phi) /Z_i,
\label{Pmin}
\end{eqnarray}
and that for depositing nothing (a vacancy)
\begin{eqnarray}
P_i(0) = 1 /Z_i,
\label{Pvac}
\end{eqnarray}
with local partition sum
\begin{eqnarray}
Z_i = 2\cosh(K\sum_{<ij>} s_j ) \exp(\Phi) +1,
\label{Z}
\end{eqnarray}
with $i=1, ..., 3^{n}$.

The homogeneous deposition model is retrieved by setting the magnetic coupling $K$ equal to zero, in which case we obtain the spin-independent probabilities
\begin{eqnarray}
P^{(0)}(+) = P^{(0)}(-)= \exp( \Phi) /Z^{(0)} \equiv P/2,
\label{Phom}
\end{eqnarray}
and the probability for depositing a vacancy
\begin{eqnarray}
P^{(0)}(0) = 1 /Z^{(0)} \equiv 1-P,
\label{Pvac0}
\end{eqnarray}
with local partition sum
\begin{eqnarray}
Z^{(0)} = 2\exp(\Phi) +1.
\label{Z0}
\end{eqnarray}
This leads to the informative relation between the reduced chemical potential and the homogeneous deposition probability $P$ in the absence of interactions,
\begin{eqnarray}
\Phi = \ln \frac{P}{2(1-P)}.
\label{PhiP}
\end{eqnarray}

We now explore in detail three physically appealing regimes: the ferromagnetic, antiferromagnetic and intermediate weak interaction cases.

\section{Ferromagnetic regime: transient Euclidean and asymptotic fractal}

\begin{figure}
 \begin{center}
       \includegraphics[width=0.95\textwidth]{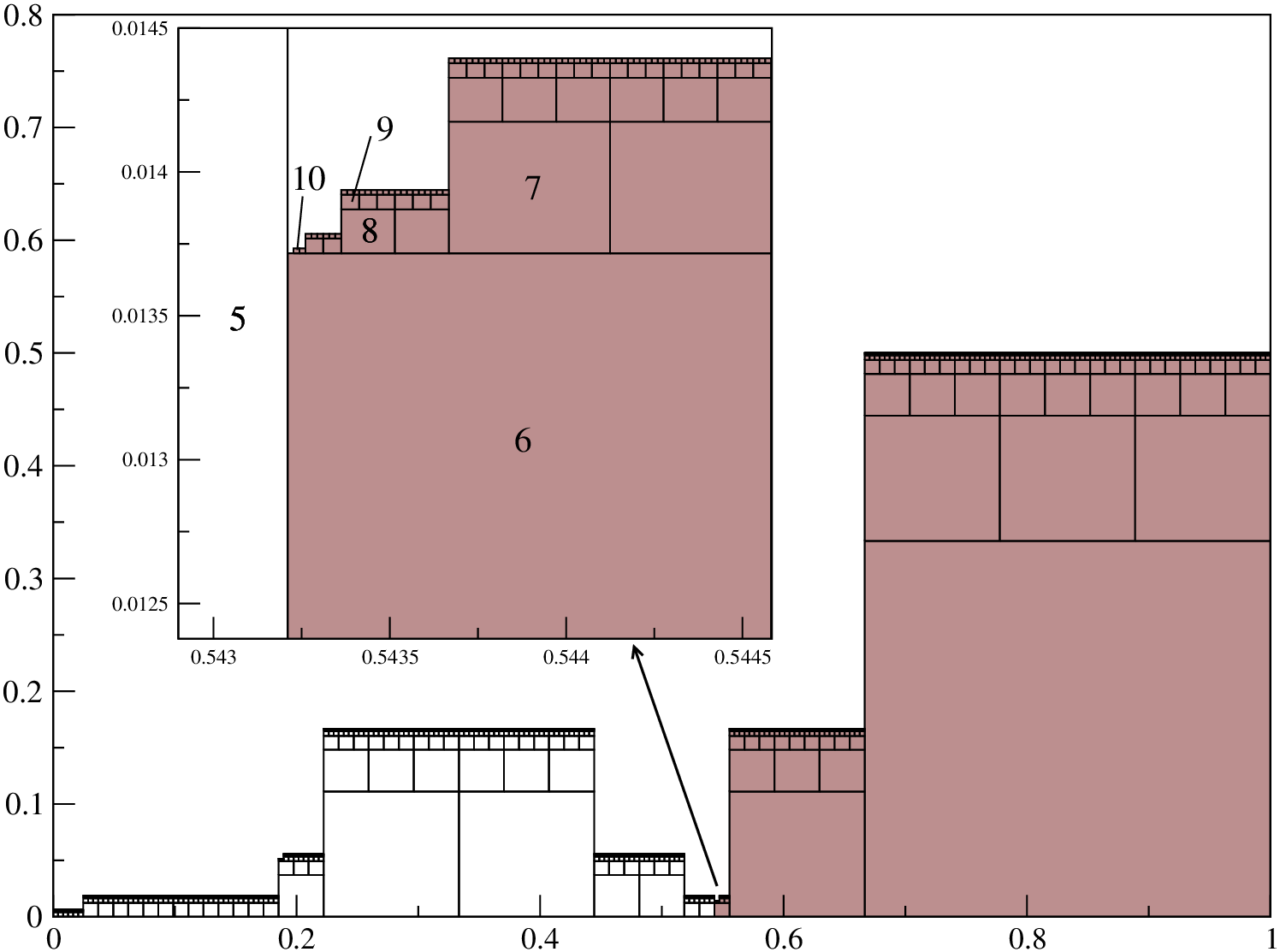}
  \end{center}
\vskip -1cm
  \emph{
\caption{Landscape for $K=5$ (strongly ferromagnetic regime) and $\Phi=-1$. Spins down are shown in dark color, while spins up are in white. The inset shows the zoomed in `fjord" area between massive domains with spins up and spins down. Deposition up to the 10th generation is shown. The numbers in the inset denote deposition generations.}\label{fjord} }
\end{figure}

\begin{figure}
 \begin{center}
       \includegraphics[width=0.99\textwidth]{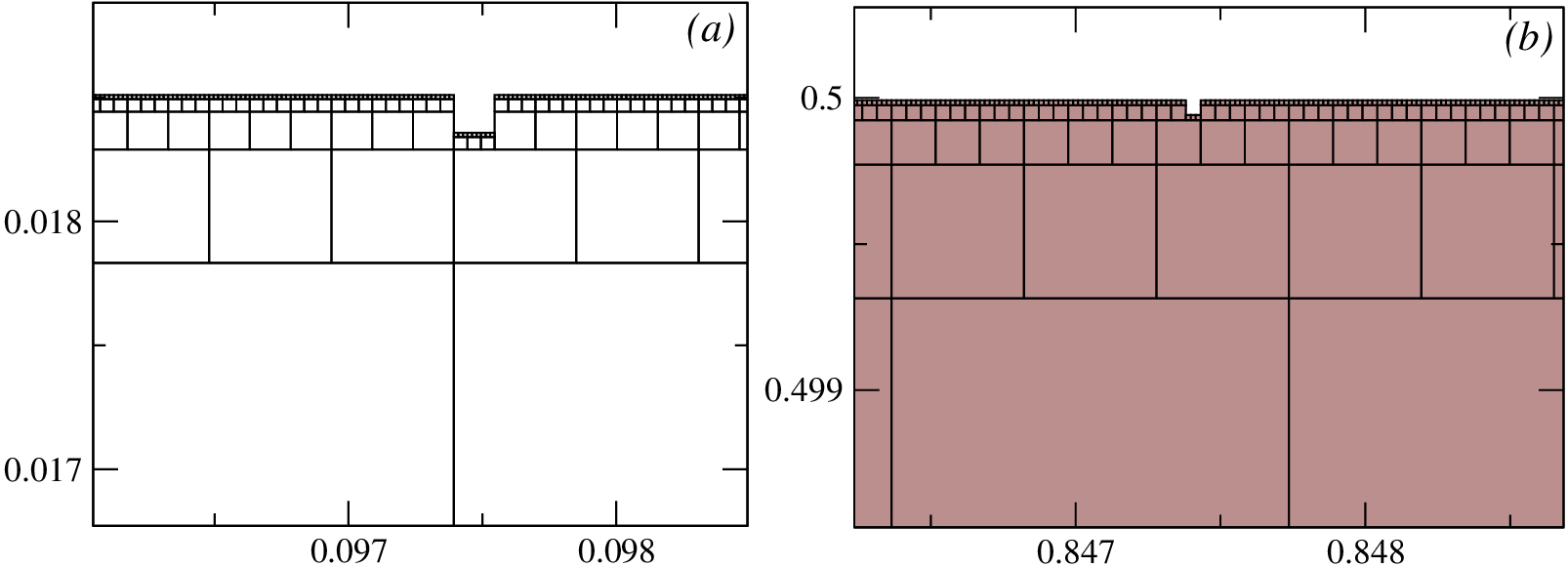}
  \end{center}
\vskip -1cm
  \emph{
\caption{Lonesome vacancies appearing in the higher deposition generations for the $K=5$ and $\Phi=-1$ landscape of Fig. \ref{fjord}. A vacancy in the 8th generation in shown in (a) and a vacancy in the 9th generation is shown in (b). The smallest blocks in this Figure belong to generation 10.}\label{holes} }
\end{figure}

Our numerical experiments indicate that in the ferromagnetic regime the landscape in the first few generations tends to partition into compact clusters of up or down spins, separated by narrow gaps or ``fjords" consisting of vacancies. An example is shown in Fig.\ref{fjord}. These compact clusters display Euclidean geometry characterized by an exponentially rapidly vanishing length increment and one might suspect that fractal growth does not occur. Surprisingly, this anticipation is false. The Euclidean behaviour turns out to be transient. The landscape evolves, for $n$ larger than some threshold value, towards a logarithmic fractal with a very small but finite length increment $\Delta L$ no matter how strongly ferromagnetic the interactions are taken. In the limit $K \rightarrow \infty$ this constant $\Delta L$ tends to zero smoothly (in a manner we will quantify), so that only in that limit a (Euclidean) behaviour different from logarithmic fractality can persist. 

Our numerical experiments clearly show a Euclidean to fractal crossover and a recovery of the logarithmic fractal for large $n$. We now proceed to prove this behaviour analytically and to derive $\Delta L$ as a function of $K$ and $\Phi$. Our basic observation, which motivates our analytic argument, is that for large positive $K$, the only features in the landscape that contribute to an increase in perimeter length are the sparse vacancies that can appear in a flat background of spins of equal sign (see Fig.\ref{holes}). These lonesome vacancies appear provided the probability for a vacancy, $P_{vac}$, times the number of attempts to put a block, being $3^n$, exceeds 1. This implies a remarkably {\em sharp crossover} from Euclidean growth to fractal growth, at the generation number $n_{\times}$ that satisfies 
\begin{eqnarray}
n_{\times} = [\ln (1/P_{vac})/\ln 3],
\label{ncrossover}
\end{eqnarray}
where the square brackets denote the smallest integer larger than the argument.
These sparse vacancies behave like a ``gas" of non-interacting particles, each contributing a length increment of $2/3^n$ in generation $n$. An obvious estimate for the length increment in generation $n$, $\Delta L_{n}$, is therefore found by multiplying the local length increment $2/3^n$ with the number of attempts $3^n$ and with the probability to place a vacancy given that the left spin {\em} and bottom spin are non-zero and equal, and the right spin is absent. This suggests
\begin{eqnarray}
\Delta L_{n} &\approx &2 P_{vac} = 2 P_i(0;++0)  = 2 /(2 \cosh(2K) \exp(\Phi)+1) \nonumber \\&\approx &2 \exp(-2K-\Phi), \,\, \mbox{for large} \; K  \mbox{and for} \,\, n \gg n_{\times}  \label{firstApprox}
\end{eqnarray}
where, for clarity, the second argument of $P_i$ shows the  spin values of the nearest neighbours of $i$. Note that the probability for a vacancy in the absence of a spin to its left, $P_i(0;0+0)$ is larger than the $P_{vac}$ we discuss here (featuring $K$ instead of $2K$ in the argument of the $\cosh$), but such a vacancy does not lead to a length increment and is therefore irrelevant. 

We compare our guess against numerical experiments. 
For example, taking $\Phi = -1$ and $K=5$, we observe Euclidean behavior up till generation 7, characterized by $\Delta L_{n} \propto 1/3^n$ (with a proportionality factor of order unity and less than 10) and then a crossover (from $n_{\times}$ = 8) to logarithmic fractal growth with $\Delta L_n$ converging to about 0.00025 from generation 12 onwards, which accurately matches our analytical estimate \eqref{firstApprox}. Note that during the Euclidean transient the proportionality factor (the number of walls) reflects the fjord topology and the boundary conditions, which are {\em open} bc's in our experiments. In contrast, during the subsequent fractal growth the walls proliferate as a constant density of vacancies develops. This density is simply proportional to $P_{vac}$. 

Interestingly, a histogram of the occurrence of vacancy probabilities teaches us that for large $K$ and large $n$ the probability $P_i(0;++0) $ results for almost all $i$. The larger probability $P_i(0;0+0)$ applies only rarely. Its occurrence is roughly equal to the number of vacancies in a given generation. This implies that the larger probability contributes little to the average vacancy probability in a given generation. To be precise, it contributes only $\exp(-K-\Phi) \times 100 \%$ to that average, which amounts to about $2\%$ in our example ($K=5$, $\Phi=-1$ and $n>12$).

This brings us to an interesting hypothesis. Since the sparse vacancies control the length increment $\Delta L$ at large $K$, deep into the ferromagnetic regime, we can describe the length increment in this regime by a random (uncorrelated) process in which the probability $P$ to put a block is just $P \approx 1- P_{vac}$. More generally, we can define $P$ as the probability for putting a block, averaged over the sites in a given generation. This amounts to a mean-field approximation for $P$, which we call $P_{mf}(n)$. We now substitute this estimate for $P$ in the exact expression for the length increment for an uncorrelated process (i.e., equivalent to a process with $K=0$), first derived in \cite{logfrac}, and obtain

\begin{eqnarray}
\Delta L_{mf}(n) = 2P_{mf}(n)(1-P_{mf}(n))/(1+P_{mf}(n)(1-P_{mf}(n))).
\label{hypo}
\end{eqnarray}

Clearly, in the regime of sparse vacancies this approximation performs very well, simply because $P_{mf}(n)$ is very close to 1, and only the factor $1-P_{mf}(n)$ matters. Also in the vicinity of $K=0$ this approximation, which is exact for $K=0$, performs very well. Therefore, we are led to compare this simple and intuitive analytical guess with our numerical results in the entire parameter space of $K$ and $\Phi$, in particular in the antiferromagnetic regime to which we turn next.

\begin{figure}
 \begin{center}
    \includegraphics[width=0.95\textwidth]{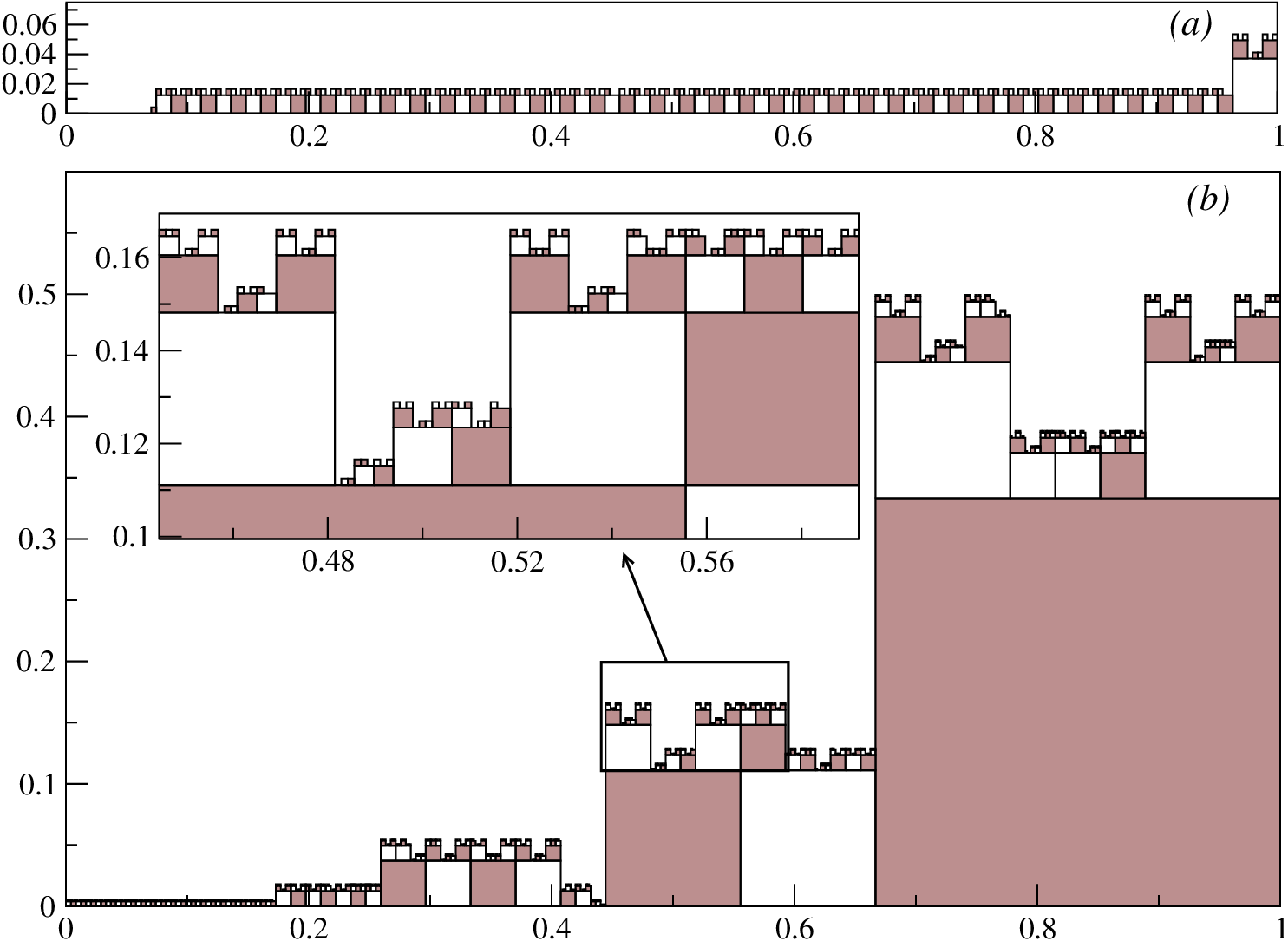}
  \end{center}
\vskip -1cm
  \emph{
\caption{``Moroccan city wall" landscapes for the strongly antiferromagnetic regime $K=-10$ for two different values of $\Phi$: (a) $\Phi=-5$ and the landscape is a result of 5 deposition generations, (b) $\Phi=-2$ and the landscape is a result of 6 deposition generations.  Part of the plot in (b) is zoomed in, so that the details of the deposition in the 6th generation can be seen better.}\label{Morroco}}
\end{figure}

\begin{figure}
 \begin{center}
    \includegraphics[width=0.95\textwidth]{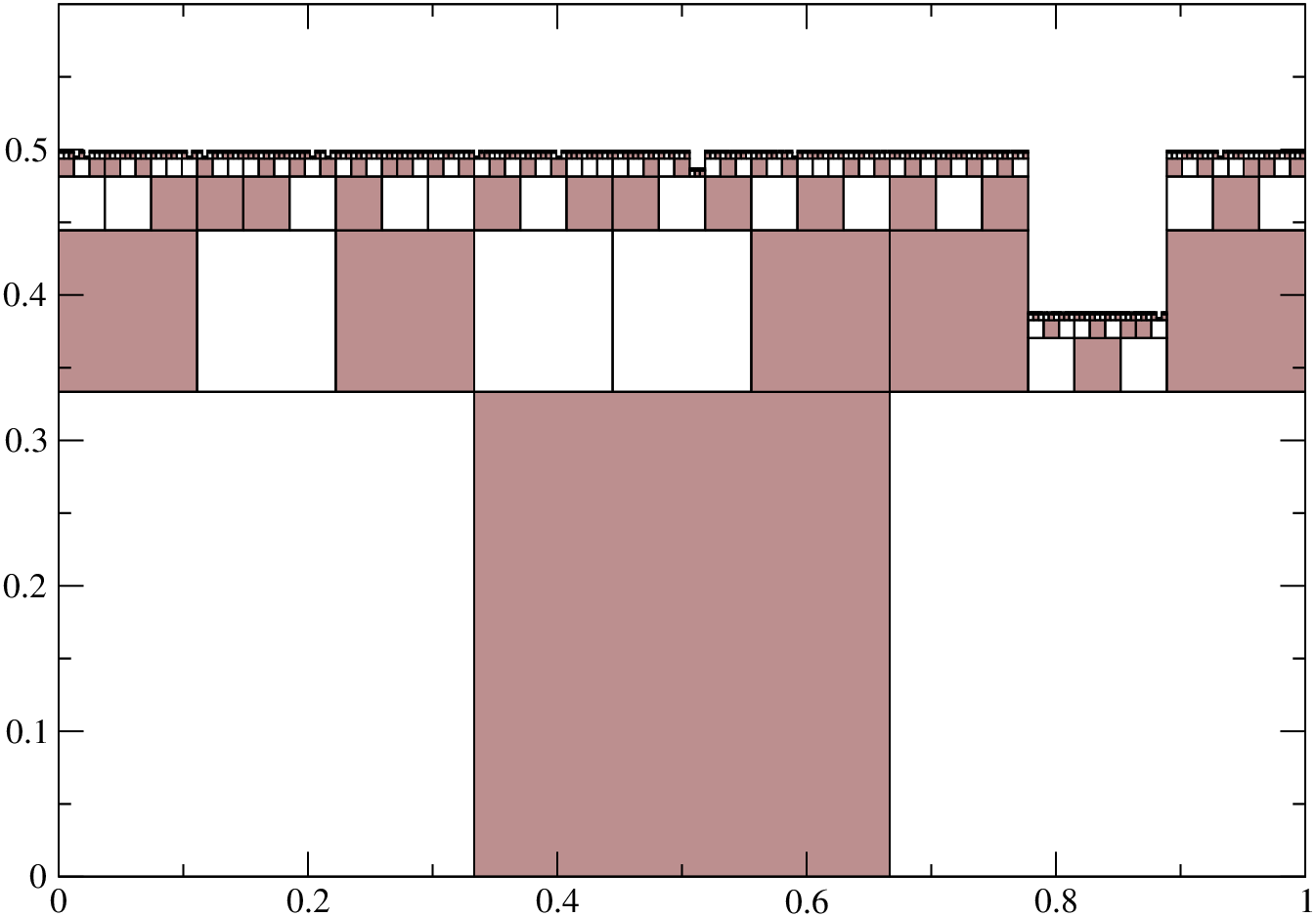}
  \end{center}
\vskip -1cm
  \emph{
\caption{An example of a landscape after 6 deposition generations in the case of strongly antiferromagnetic correlations ($K=-5$ and  $\Phi=2$).}\label{AFMpos_phi}}
\end{figure}

\begin{figure}
 \begin{center}
 \includegraphics[width=0.99\textwidth]{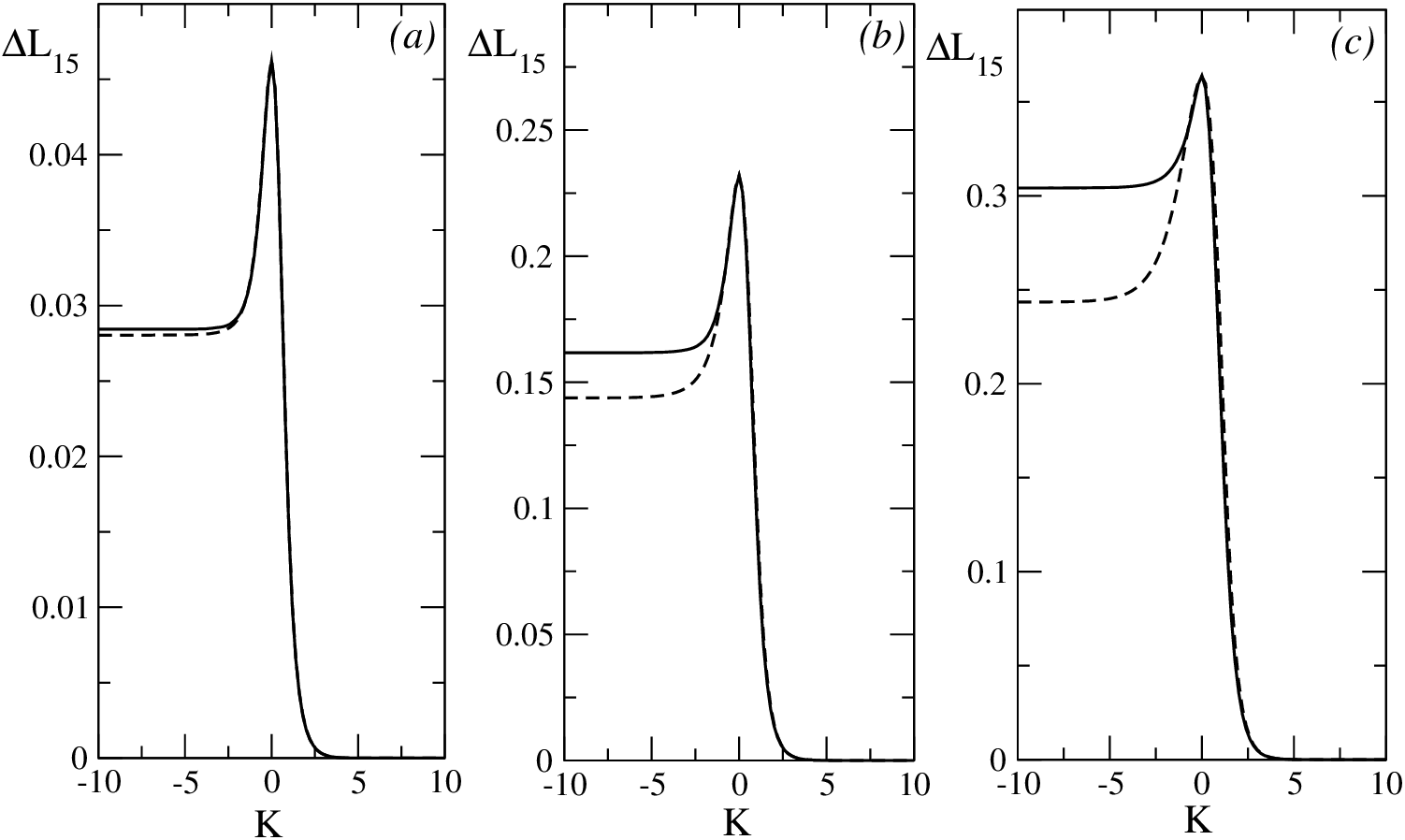}
 \end{center}
\vskip -1cm
  \emph{
\caption{Comparison of numerical results for the length increment $\Delta L_{15}$ versus interaction strength $K$ with the theoretical prediction of Eq. \eqref{hypo} for different values of $\Phi$: (a) $\Phi=3$;  (b) $\Phi=1$, (c) $\Phi=0$. Solid lines correspond to the numerical simulation and dashed curves correspond to the theoretical estimate of Eq. \eqref{hypo}.  }
\label{posphi}}
\end{figure}

\begin{figure}
 \begin{center}
 \includegraphics[width=0.99\textwidth]{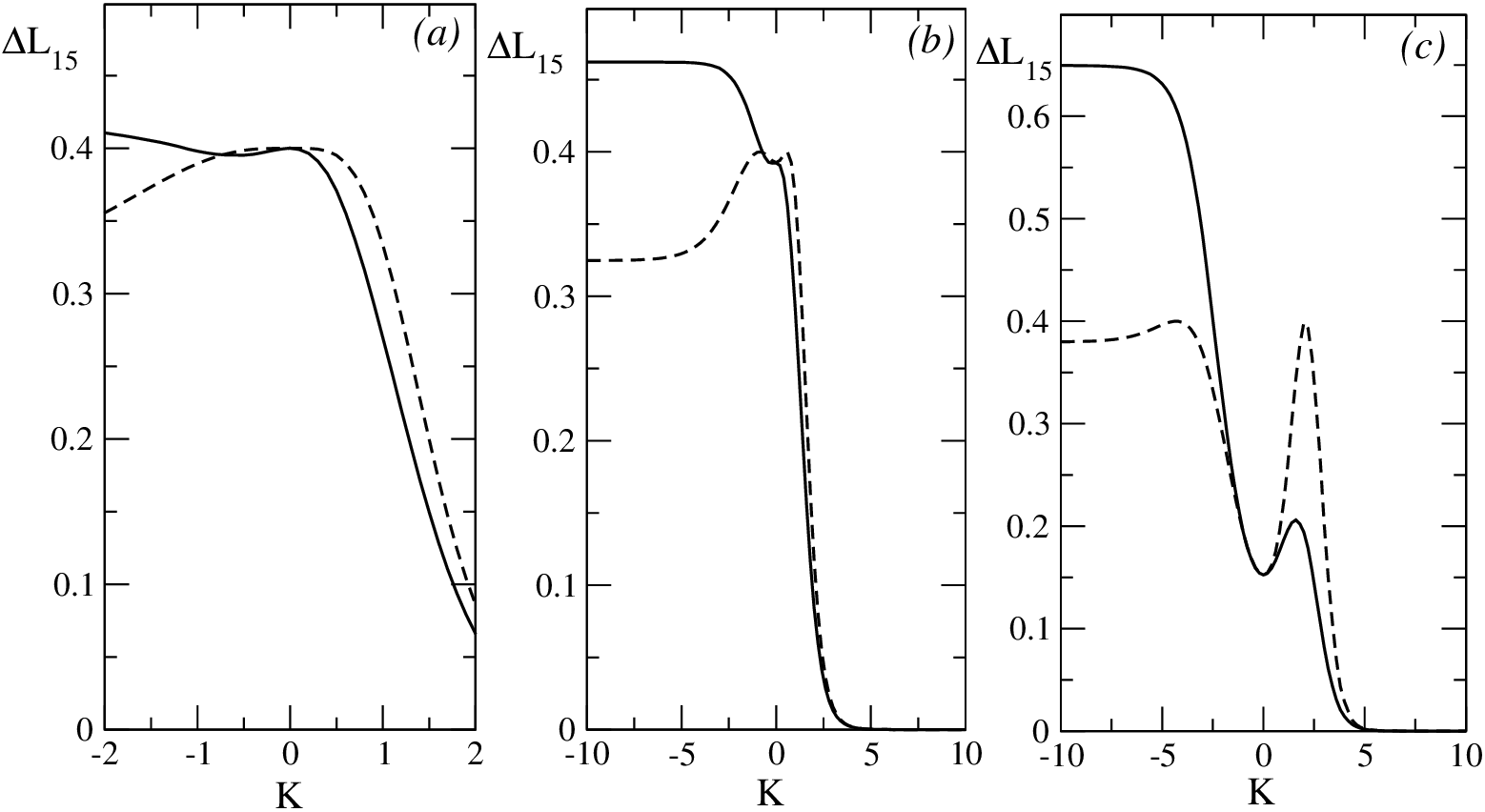}
 \end{center}
\vskip -1cm
  \emph{
\caption{Comparison of numerical results for the length increment $\Delta L_{15}$ versus interaction strength $K$ with the theoretical prediction in Eq. \eqref{hypo} for different values of $\Phi$: (a) $\Phi=-ln2$;  (b) $\Phi=-1$, (c) $\Phi=-3$. Solid lines correspond to the numerical simulation and dashed curves correspond to the theoretical estimate of Eq. \eqref{hypo}. }
\label{negphi}}
\end{figure}

\section{Antiferromagnetic regime: fractality controlled by vacancies and spin frustration}
Deep into the antiferromagnetic regime, $K<0$ and $\mid K\mid $ large, the physics of the deposition is governed by anticorrelations between spins. We can distinguish two main types of behaviour, depending on the value of the reduced chemical potential $\Phi$. When vacancies are promoted, i.e., at large negative values of $\Phi$, e.g. $\Phi =-5$, we see an ordered interplay of anticorrelation and vacancy formation.  The resulting geometry resembles the pattern occurring in Moroccan city walls (see Fig.\ref{Morroco}). On the other hand, when vacancies are suppressed, through positive values of $\Phi$, we see anticorrelation and frustration of spins accompanied by sparse vacancies as illustrated in Fig.\ref{AFMpos_phi}. 

These observations prompt us to test our hypothesis $\eqref{hypo}$ against numerical results. For this purpose we provide, in Figs.\ref{posphi} and \ref{negphi}, the length increment obtained in the highest generation that we can deal with computationally, $n=15$, versus $K$ at a fixed value of $\Phi$. We cover the entire range of interaction, from strongly antiferromagnetic to strongly ferromagnetic. 

We notice the following trends. When vacancies are suppressed ($\Phi >0$, as in Fig.\ref{posphi}), our analytic hypothesis performs very well throughout the entire range of interaction. In contrast, when vacancies are more abundant ($\Phi <0$, as in Fig.\ref{negphi}), density fluctuations render the mean-field approximation less precise. The antiferromagnetic regime, which features spin frustration effects that promote vacancies, is more sensitive to this than the ferromagnetic one. Note that in all cases a logarithmic fractal behaviour is found, although for large negative values of $\Phi$ the convergence of the length increment versus $n$ becomes poor. Our precise investigation is hereby limited to a range of $\Phi$ that corresponds roughly to $\Phi > -3$.

The antiferromagnetic regime holds an interesting surprise. When vacancies are sparse, the length increment deep in the antiferromagnetic regime can be calculated exactly. The result is 
\begin{eqnarray}
\Delta L_{\infty} = 2(\sqrt{13}-3)/(2 \exp(\Phi) +1).
\label{exactfrus}
\end{eqnarray}
This can be proven as follows. The local probability for placing a vacancy  takes one out of the following four values: $1/(2\cosh(mK)\exp(\Phi)+1)$, with $m \in \{0,1,2,3\}$. Note that the case $m=0$ corresponds to a {\em frustrated spin}, the two neigbours of which are of opposite sign. Examining the degeneracies, or frequencies of occurrence, of these values in a histogram, reveals that, for large $\mid K \mid$, the case $m=0$ occurs on roughly $60\%$ of the sites and the case $m=2$ on roughly $40\%$ of the sites. The occurrence of the remaining cases is negligible. Now, since the probability for placing a vacancy is very small for $m=2$ as compared to $m=0$, only frustrated spins can give rise to vacancy formation and thus the average probability for putting a vacancy is very close to $P_{vac} = 60\% \times 
1/(2\exp(\Phi)+1)$. In order to estimate the coefficient accurately, we have recourse to a reasoning based on conditional probabilities for obtaining frustrated spins. Our goal in this exercise is to calculate the probability $P_{frus}$ that a spin is frustrated, i.e., that it has precisely two neighbouring spins that are of opposite sign.

We logically deduce that, in a given generation, a frustrated spin is followed, to its right, by a frustrated spin with probability $1/2$. Further, a non-frustrated spin at $i$ is followed, to its right, by a frustrated spin at $i+1$ {\em provided} it rests on the same block (put in the previous generation) as the spin at $i+1$. The probability that $i$ and $i+1$ are on the same block, is $2/3$. On the other hand, a non-frustrated spin at $i$ is followed, to its right, by a frustrated spin at $i+1$ with probability $P_{frus}$ {\em provided} it rests on a different block (put in the previous generation) than the spin at $i+1$. The probability that $i$ and $i+1$ are on different blocks, is $1/3$, and the probability $P_{frus}$ invoked here is the probability that the rightmost underlying block spin was frustrated in the previous generation. 

This reasoning leads us to the following self-consistency requirement, based on conditional probability calculus,
\begin{eqnarray}
P_{frus}= P_{frus}/2+(2/3+P_{frus}/6)(1-P_{frus}),
\label{selffrus}
\end{eqnarray}
from which we obtain the exact result
\begin{eqnarray}
P_{frus}= \sqrt{13}-3 = 0.605551...,
\label{selffrus}
\end{eqnarray}
which explains why we see a close to $60\%$ incidence of frustrated spins in the numerical experiments. In fact, our simulations are sufficiently accurate to detect this coefficient with an error of less than $0.5\%$. For the length increment deep in the antiferromagnetic regime with thus obtain
\eqref{exactfrus}, since in generation $n$ each vacancy contributes a local length increment of $2/3^n$ and there are $3^n$ attempts.

\section{Intermediate weak-interaction regime near $K=0$: local extrema of $\Delta L  (K)$}
The regime near $K=0$ is, in spite of its seeming simplicity, very interesting. The anchoring point, $K=0$, where the mean-field approximation $\eqref{hypo}$ is exact, is flanked by regions that display remarkable behaviour, when we inspect Figs.\ref{posphi} and \ref{negphi}. Firstly, when vacancies are promoted ($\Phi <0$) the length increment displays a minimum at $K=0$. This minimum reflects that the probability $P$ for placing a block, which typically goes through a local minimum at $K=0$ due to the absence of interaction, is smaller than $1/2$. Indeed, in this range of $P$, the length increment is an increasing function of $P$ \cite{logfrac}.

Conversely, for values of the chemical potential that suppress vacancies, the value of $P$ at $K=0$ is larger than 1/2. In that range, a local minimum of the probability for placing a block, typically occurring at $K=0$, will lead to a local maximum of the length increment, because the exact $\Delta L$ at $K=0$ is a decreasing function of $P$ for $P>1/2$ \cite{logfrac}. This explains why $K=0$ can be associated with a minimum or a maximum of the length increment, depending on the chemical potential. The cross-over between a local minimum of the length increment at $K=0$ and a local maximum at $K=0$, occurs roughly for the value of the chemical potential that renders $P=1/2$ in \eqref{PhiP}, which entails $\Phi = -\ln 2 = -0.693...$. At about this value of the chemical potential we ought to see an inflection point in $\Delta L$ versus $K$, at $K=0$ (see Fig. \ref{negphi}({\it a})).

Incidentally, it is easy to see that the probability for placing a block is minimal at $K=0$. Indeed, from \eqref{Pvac} and \eqref{Z} it is conspicious that the probability for putting a vacancy decreases when $K$ is moved away from zero, since $\cosh(x) \geq 1$. This signifies that switching on the interaction, ferromagnetic or antiferromagnetic, acts so as to increase the probability for placing a block. 

Further, an auxiliary local maximum can occur, tycally in the ferromagnetic regime $K>0$. This is perspicuous in some of our curves, notably those with $\Phi <0$. Its explanation is as follows. As we increase $K$ from $K=0$, the average probability for putting a block increases from a value below 1/2 towards values close to 1 for large $K$. In doing so, the value of $P_{mf}$ inevitably passes through 1/2, which is where $\Delta L_{mf}$ has a maximum as a function of $P_{mf}$. Therefore, this local maximum will also be reflected in the length increment as a function of $K$, as long as the mean-field approximation is reliable.

\begin{figure}
 \begin{center}
 \includegraphics[width=0.95\textwidth]{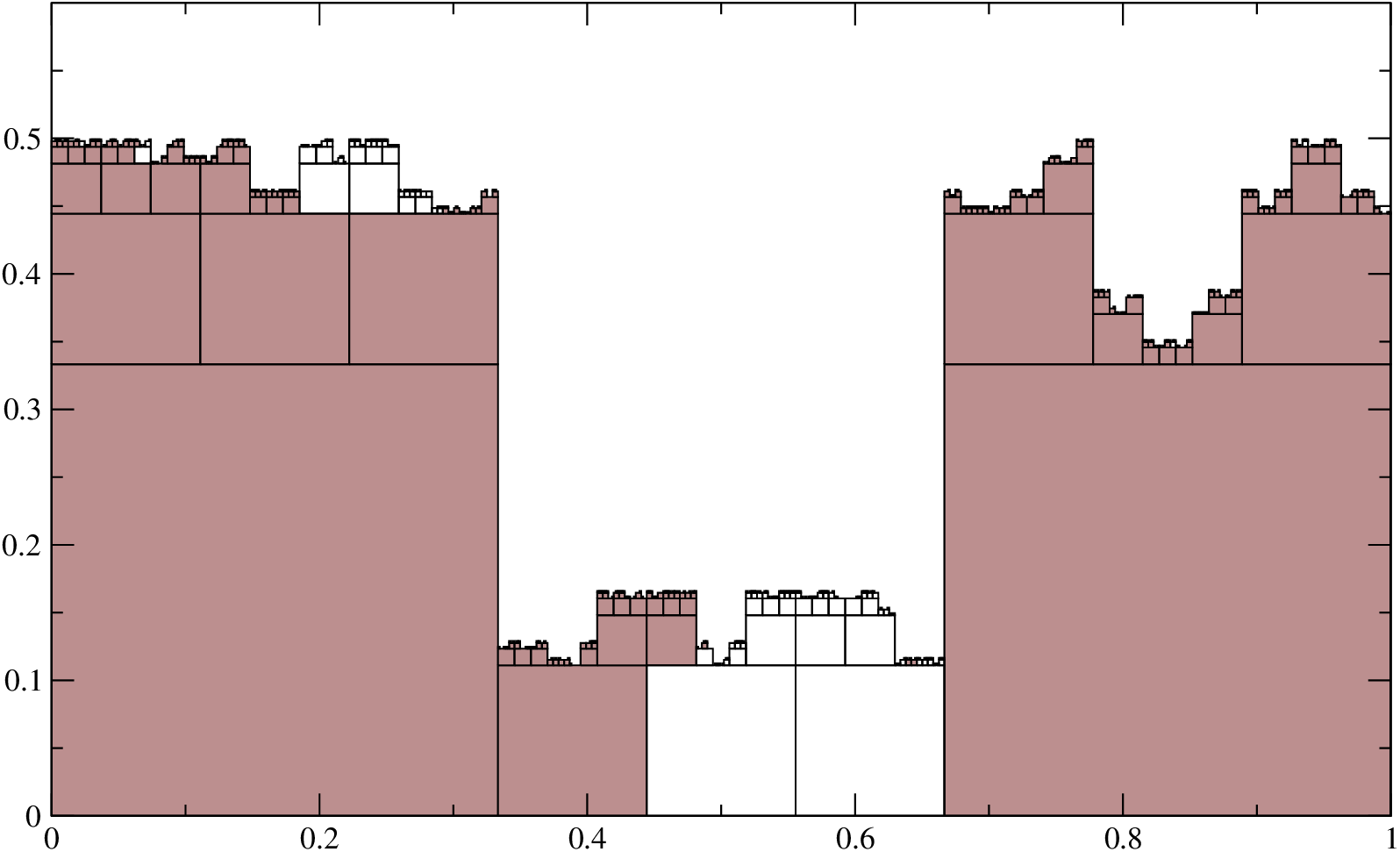}
 \end{center}
\vskip -1cm
  \emph{
\caption{Landscape after 6 deposition generations for the weakly ferromagnetic regime $K=1, \Phi=-1$.}
\label{weak_afm}}
\end{figure}

\begin{figure}
 \begin{center}
 \includegraphics[width=0.95\textwidth]{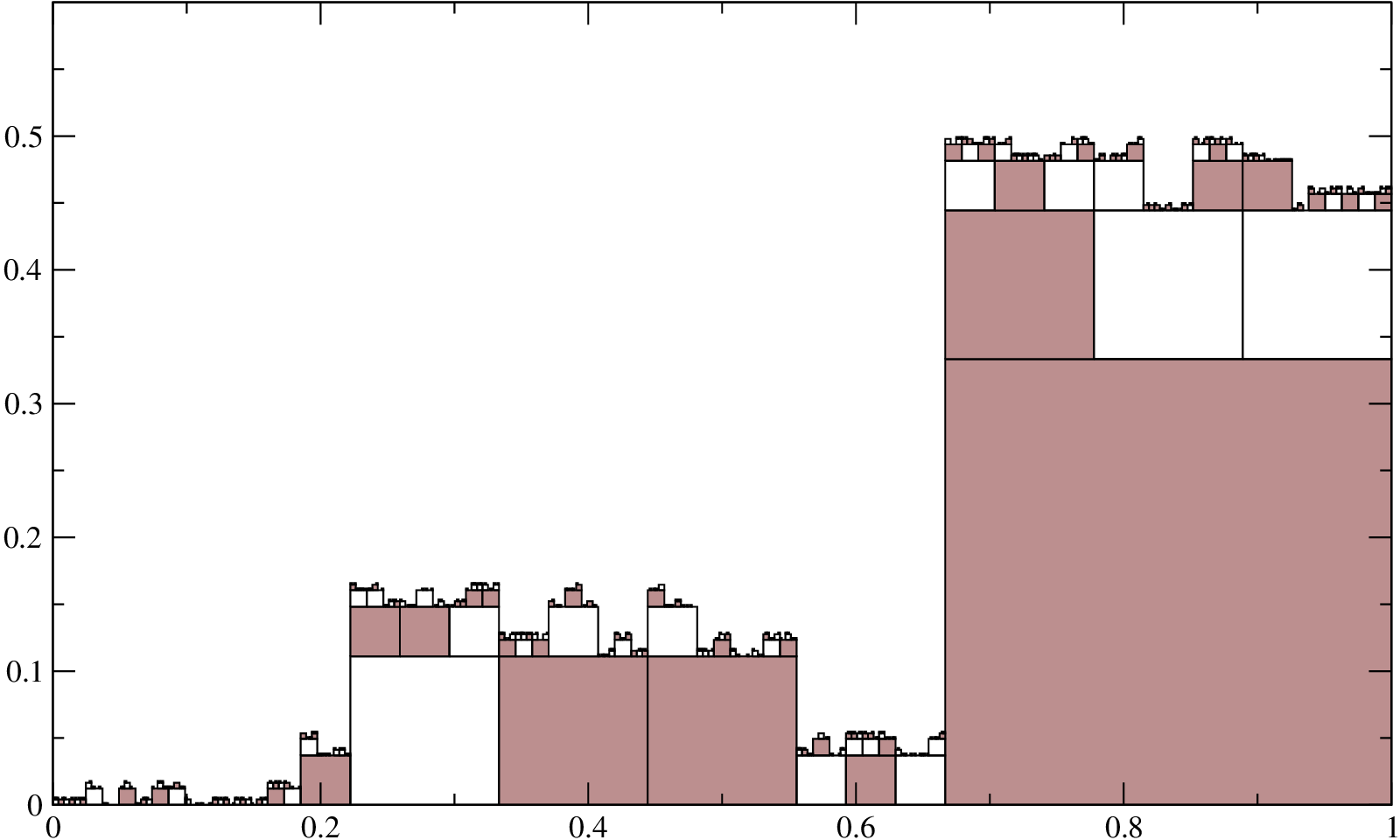}
 \end{center}
\vskip -1cm
  \emph{
\caption{Landscape after 6 deposition generations for the weakly antiferromagnetic regime $K=-1, \Phi=-1$.}
\label{weak_fm}}
\end{figure}

Typical landscapes for weakly antiferromagneitc ($K=-1$) and weakly ferromagnetic regimes ($K=1$) and for fixed $\Phi=-1$ are displayed in Figs. \ref{weak_afm} and \ref{weak_fm}. 

\section{Conclusions}
The main purpose of this paper has been to investigate how a class of fractal surfaces, the so called logarithmic fractals characterized by a constant length increment in every generation, respond when correlations are introduced that interfere with the random character of the hierarchical deposition from which they are grown. Exploring the simplest way of introducing correlations, through the attribution of magnetic degrees of freedom or ``spins" to deposited blocks, and allowing for ferro- or antiferromagnetic interactions between them, we have uncovered remarkable behaviour. 

Our numerical experiments are supported by theoretical predictions in the form of analytical results pertaining to the statistical mechanical properties of the model. They clearly demonstrate that, while magnetic interactions can have a strong smoothing effect on the surface of the deposited material (in case of ferromagnetic couplings) or a strong roughening effect (in case of antiferromagnetic couplings), the logarithmic fractal character is preserved throughout. Even deep into the ferromagnetic regime, where massive aggregation of blocks with aligned spins occurs, the logarithmic fractal character recovers asymptotically, after a Euclidean transient. In this regime the amplitude of the logarithmic fractal is greatly reduced with respect to the model without interactions and at the same chemical potential governing the block deposition probability. We have calculated the reduced amplitude analytically. 

Another, distinct, manifestation of the robustness of logarithmic fractals has been revealed in the strongly antiferromagnetic regime, governed by anticorrelation of spins and by spin frustration, occurring when a deposited block has two neighbours of opposite spin. In that regime we found in our numerical experiments, and proved in our analytic scrutiny, that the vacancies arising from frustrated spins are responsible for the main contribution to the surface roughness. Again, the logarithmic fractal character is stable, and its amplitude can be obtained exactly, in the limit of sparse vacancies.

In general, we have found that a mean-field approximation based on the average probability $P_{mf}$ to place a block in a given generation of deposition, provides a reliable tool for estimating the asymptotic length increment $\Delta L$ of the landscape. A good approximation for the latter is obtained when $P_{mf}$ is used as input in the exact expression of $\Delta L$ for the {\em uncorrelated} model, derived in the pioneering paper \cite{logfrac}. The quality of this mean-field approach deteriorates progressively as more and more vacancies appear, by manipulating the chemical potential for the blocks.

Finally, in the weak interaction regime, our simulations and calculations unveil that the non-interacting model is {\em extremal} in the sense that it corresponds to a minimum (or maximum) of the amplitude of the logarithmic fractal as a function of the magnetic coupling. The character of this extremum is easily predicted from the properties of the uncorrelated model. 

In sum, we have demonstrated the robustness of a particular random and hierarchical growth law for rough surfaces with respect to the introduction of spatial correlations. The correlations have been modeled through magnetic interactions between pseudo-spins. The extension of this model to higher dimensions (in particular to two-dimensional substrates) is immediate, as it was for the uncorrelated model \cite{several}. Also the generalization to other boundary conditions and substrate conditions is straightforward and not expected to change our results qualitatively. The magnetic hierarchical deposition model complements the magnetic Eden model and its variants physically and fundamentally in that it provides an alternative paradigm for tunable surface roughness.

{\em Acknowledgements} \\
We gratefully acknowledge KU Leuven Research Grant OT/11/063.

\end{document}